\documentclass[prb,twocolumn,showpacs,superscriptaddress,groupedaddress,amsmath,amssymb]{revtex4-1}
\usepackage{graphicx}
\usepackage{dcolumn}
\usepackage{bm}
\usepackage{hyperref}
\usepackage{color}
\usepackage{ulem} 

\begin{document}

\title{
Large-Tiny Spin-Stripe SDW Order in La$_3$Ni$_2$O$_7$ via $^{139}$La-NQR Spectroscopy: Reconsideration Based on the $Am2m$ Structure
}

\author{
Mitsuharu Yashima$^{1}$
}
\email{yashima.mitsuharu.es@osaka-u.ac.jp}
\author{\hspace{-1.2em}
, Hongnam Lee$^1$,
Takumi Ino$^1$,
Nina Seto$^1$,
Masataka Kakoi$^{1,2}$,\\
Hiroya Sakurai$^3$,
Yoshihiko Takano$^{3,4}$, 
and Hidekazu Mukuda$^1$
}

\affiliation{$^1$Graduate School of Engineering Science, The University of Osaka, Toyonaka, Osaka 560-8531, Japan} 
\affiliation{$^2$Department of Physics, The University of Osaka, Toyonaka, Osaka 560-0043, Japan} 
\affiliation{$^3$National Institute for Materials Science (NIMS), Tsukuba, Ibaraki 305-0047, Japan} 
\affiliation{$^4$Graduate School of Pure and Applied Sciences, University of Tsukuba, Tsukuba, Ibaraki 305-8577, Japan}

\date{\today}

\begin{abstract}
The lower-symmetry $Am2m$ structure was recently proposed as the intrinsic ambient-pressure phase of La$_3$Ni$_2$O$_7$ based on synchrotron x-ray diffraction, whereas earlier studies investigated its properties assuming the commonly accepted $Amam$ structure.
Here, we reexamine the crystal structure in La$_3$Ni$_2$O$_7$ and the spin configuration of spin-density-wave (SDW) state via $^{139}$La nuclear quadrupole resonance (NQR). 
The observation of two distinct La(2) sites in the NQR spectra is naturally accounted for by the $Am2m$ structure without any special assumptions, whereas it cannot be reconciled within the $Amam$ framework.
Within the $Am2m$ framework, we reanalyze the spin configuration of the SDW phase and identify it as a large-tiny spin-stripe structure, arising naturally from two inequivalent Ni sites with large and tiny magnetic moments.
These results establish the itinerant nature of the SDW order at ambient pressure in La$_3$Ni$_2$O$_7$, providing a basis for understanding its relationship to the emergence of high-$T_c$ superconductivity under high pressure.
\end{abstract}

\maketitle

To elucidate the mechanism of high-$T_{c}$ superconductivity emerging in the bilayer nickelate La$_3$Ni$_2$O$_7$ (La327) under high pressure~\cite{Sun_2023}, it is essential to clarify the nature of the density-wave (DW) phase associated with spin- and/or charge-density-wave (SDW/CDW) order at ambient pressure~\cite{Wu, Liu2023, Liu2024, Hou_replication_2023, Yanan-Zhang_replication, G-Wang_2024, Wang_La2PrNi2O7,Chen_RIXS, Taniguchi, Kakoi, Meng, Chen_muSR, Khasanov, Ren, Gupta, Zhao_NMR,Yashima327,Plokhikh,Luo,Lee,Chen-single-neutron}.
Several possible candidates have been proposed, including a single spin-charge (spinless) stripe order~\cite{Chen_RIXS,Chen_muSR,Khasanov,Ren,Yashima327,Plokhikh,Lee,Chen-single-neutron}, a double spin-stripe order~\cite{Chen_RIXS,Gupta,Zhao_NMR}, and a double charge-stripe order~\cite{Luo}. 
Despite extensive studies, the nature and ordering pattern of the DW phase remain unclear.
In addition, the ambient-pressure crystal structure is also under debate.
The orthorhombic $Amam$ structure (space group No.~63) has long been regarded as the ambient-pressure crystal structure~\cite{Sun_2023}, while some reports have suggested the $Fmmm$ structure (No.~69)~\cite{Neutron,Fmmm-Zhang,Zhang_JSSC2026}.
Furthermore, recent synchrotron x-ray diffraction (SXRD) measurements, enabled by a high photon flux and a large dynamic range, have proposed a lower-symmetry $Am2m$ structure (No.~38) for high-quality single crystals~\cite{Misawa}.
In the $Amam$ and $Fmmm$ structures, which feature a single distinct Ni site, the NiO$_6$ octahedra are tilted from the $c$ axis toward the $b$ axis for $Amam$ and point straight along the $c$ axis for $Fmmm$. In contrast, the $Am2m$ structure incorporates two crystallographically inequivalent Ni sites with different local states, characterized by tilted NiO$_6$ octahedra with distinct octahedral volumes.

Here, we reexamine the $^{139}$La nuclear quadrupole resonance (NQR) spectra of La$_3$Ni$_2$O$_{7.0}$ to identify the intrinsic local structures.
Nuclei with spin $I \ge 1$ possess a nuclear quadrupole moment $Q$, which interacts with the local electric field gradient (EFG) and serves as a highly sensitive probe of the local crystal structure.
This interaction is described by the nuclear quadrupole Hamiltonian,
$\mathcal{H}_Q = \frac{h\nu_Q}{6} [3I_z^2 - I(I+1) + \frac{\eta}{2} (I_+^2+I_-^2)]$,
where $\nu_Q=3eQV_{zz}/[2I(2I-1)h]$ is the NQR frequency, $V_{zz}$ is the principal component of the EFG tensor, and $\eta=|V_{xx}-V_{yy}|/V_{zz}$ is the EFG asymmetry parameter.
For $^{139}$La ($I = 7/2$), three NQR transitions [$1\nu_Q$($\pm\frac{1}{2}$$_{ }$$\leftrightarrow$$\pm\frac{3}{2}$), $2\nu_Q$($\pm\frac{3}{2}$$_{ }$$\leftrightarrow$$\pm\frac{5}{2}$), and $3\nu_Q$($\pm\frac{5}{2}$$_{ }$$\leftrightarrow$$\pm\frac{7}{2}$)] are expected for each La site.
In La327, there are two crystallographically inequivalent La sites: La(1), located between the NiO$_2$ planes, and La(2), located outside the NiO$_2$ bilayers.
These sites are readily distinguished in the NMR/NQR spectra by their different $\nu_Q$ values~\cite{Fukamachi_2001,Kakoi,Zhao_NMR,Yashima327,Wang_La2PrNi2O7,Luo}.
Further analysis of the $^{139}$La(2)-NQR spectra revealed at least two distinct La(2) sites, denoted La(2)$_{\rm a}$ and La(2)$_{\rm b}$, even in high-quality samples~\cite{Yashima327}, whereas only a single La(2) site is expected for an ideal $Amam$ structure. 
Thus, previous $^{139}$La-NQR studies of La327 assigned the narrow La(2)$_{\rm a}$ peak to the ideal La(2) site and attributed the broader La(2)$_{\rm b}$ peak to possible defect-related origins, such as (i) interfaces associated with La$_4$Ni$_3$O$_{10}$ (La4310) stacking faults~\cite{Wang_La2PrNi2O7,Luo}, (ii) contamination by the $Fmmm$ phase~\cite{Zhao_NMR}, and (iii) inner apical oxygen vacancies~\cite{Yashima327}.
If the recently proposed $Am2m$ structure is indeed intrinsic, these interpretations should be reconsidered within the $Am2m$ framework.

\begin{table}[!b]
\caption{
Comparison of NQR parameters ($\nu_{Q}$, $\eta$) for La(2)$_{\rm a}$ and La(2)$_{\rm b}$ obtained from NQR experiments at 160 K with DFT+$U$ calculations for $Am2m$, $Amam$, and $Fmmm$. The $Am2m$ model reproduces the two La(2)$_{\rm a,b}$ sites and shows good agreement with the experimental values.
The calculated values for $Fmmm$ are close to those reported previously~\cite{Zhao_NMR}.
}
    \label{NQR-Table}
    \begin{ruledtabular}
    \renewcommand{\arraystretch}{1.2}
    \begin{tabular}{lcccc}
    & $\nu_Q^{\rm a}$(MHz) & $\eta^{\rm a}$ &  $\nu_Q^{\rm b}$(MHz) & $\eta^{\rm b}$ \\
    \hline
    Experiment & 5.913 & 0.025   & 5.383 & 0.153   \\
    \hline
    Cal. ($Am2m$)   & 5.911 & 0.014  &  5.279 & 0.135  \\  
    Cal. ($Amam$)   & 5.757 & 0.028 & -- & --    \\  
    Cal. ($Fmmm$)   & -- & --   & 5.016 & 0.136   \\  
    \end{tabular}
    \end{ruledtabular}
\end{table}

In this Letter, we show that the La(2)$_{\rm a}$ and La(2)$_{\rm b}$ sites observed in the $^{139}$La(2)-NQR spectra of La$_3$Ni$_2$O$_{7.0}$~\cite{Yashima327} can be consistently accounted for by the NQR parameters calculated for the $Am2m$ structure~\cite{Misawa}.
In the SDW phase, the tiny and large internal fields ($H_{\rm int}$) at La(2)$_{\rm a}$ and La(2)$_{\rm b}$ are derived from the Ni$_{\rm a}$ (tiny-spin) and Ni$_{\rm b}$ (large-spin) channels along the $a$ axis, respectively.
These results establish that high-quality La$_3$Ni$_2$O$_{7.0}$ adopts the lower-symmetry $Am2m$ structure and exhibits a commensurate SDW order with a large-tiny spin-stripe pattern below $T_{\rm SDW}=150$ K at ambient pressure.

The polycrystalline La$_3$Ni$_2$O$_{7+\delta}$ ($\delta \sim 0.0$) sample was prepared by the solid-state reaction method described elsewhere~\cite{Ueki,Sakurai}.
The X-ray diffraction confirmed a single phase of the orthorhombic structure with lattice parameters of $a$ = 5.3920~\AA, $b$ = 5.4510~\AA, and $c$ = 20.533~\AA~\cite{Ueki}.
The oxygen deficiency $\delta$ was determined by thermogravimetric analysis.
The $^{139}$La(2)-NQR spectrum was obtained at zero external field using a coarse powder sample. 
The NQR parameters, $\nu_Q^{\rm cal}$ and $\eta^{\rm cal}$, for each La site were calculated using the density functional theory plus Hubbard $U$ (DFT+$U$) technique as implemented in the \textsc{Quantum Espresso} package~\cite{QE}.
The exchange-correlation energy was evaluated within the generalized gradient approximation parametrized by the Perdew-Burke-Ernzerhof (PBE) functional~\cite{Perdew}, based on the experimental lattice parameters and atomic coordinates for $Amam$~\cite{Ueki,Ling}, $Fmmm$~\cite{Fmmm-Zhang}, and $Am2m$~\cite{Misawa}.
An effective Hubbard $U$ of 3.5 eV, as a typical value within the range of 3 -- 4 eV~\cite{Sun_2023,Zhao_NMR,Sakakibara_PRL,Yang}, was applied to the Ni-$3d$ orbitals.
The calculated NQR parameters are summarized in Table \ref{NQR-Table}.

\begin{figure}[!t]
\hspace*{-0.3cm}
\includegraphics[width=8.0cm]{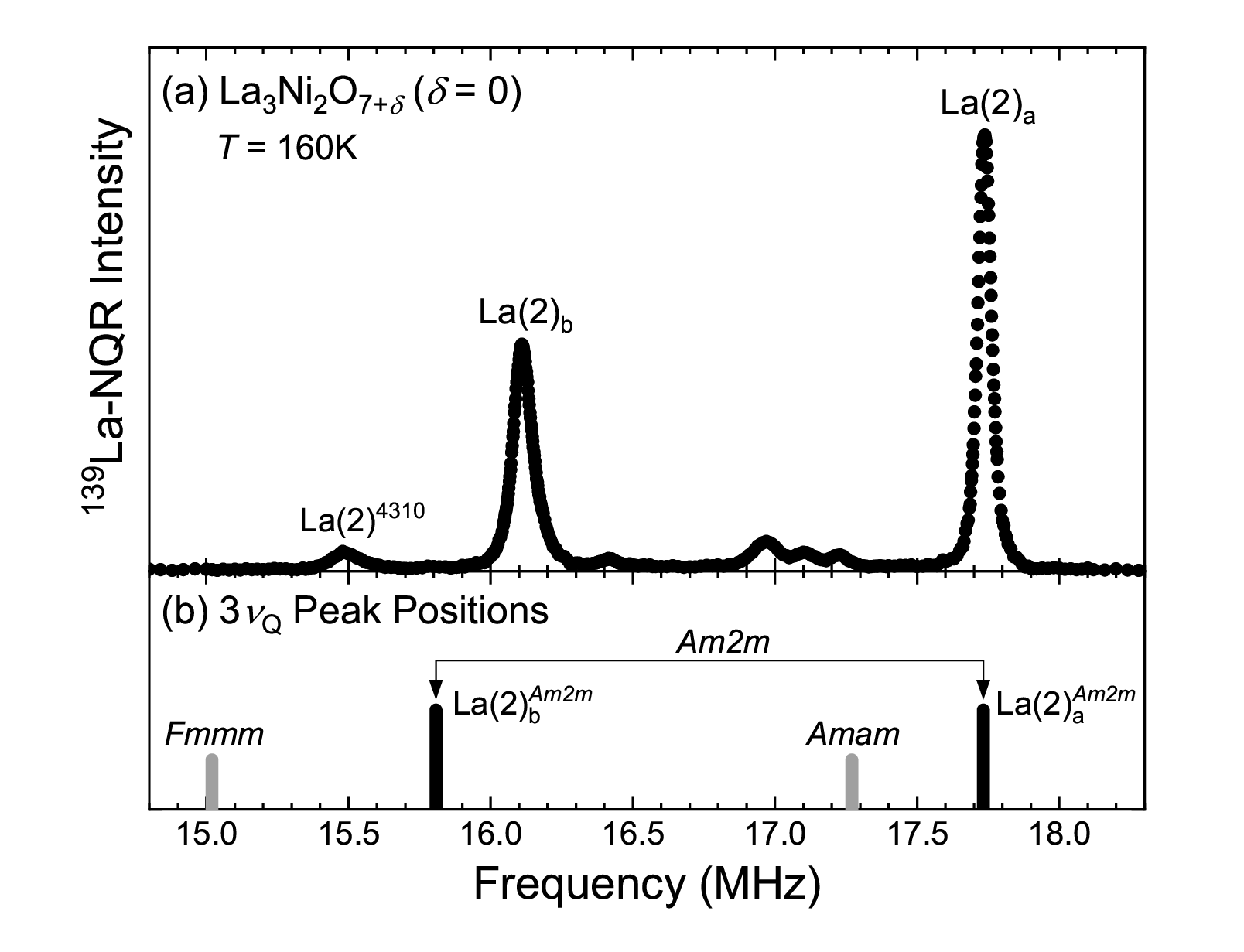}
\vspace*{-0.4cm}
\caption[]{\footnotesize 
(a) $^{139}$La(2)-NQR spectrum at 160 K ($>T_{\rm SDW}$) in La$_3$Ni$_2$O$_{7.0}$. 
Despite the high quality of the sample, two peaks corresponding to La(2)$_{\rm a}$ and La(2)$_{\rm b}$ are observed with nearly equal intensities, consistent with lower-symmetric $Am2m$ rather than $Amam$.
(b) Calculated $3\nu_Q$ peak positions at the La(2) site for $Am2m$, $Amam$, and $Fmmm$ using the NQR parameters in Table \ref{NQR-Table}. 
La(2)$_{\rm a}^{Am2m}$ and La(2)$_{\rm b}^{Am2m}$ are obtained by calculation based on $Am2m$, which are in good agreement with the experimental results.
In (a), the peaks around 16.9--17.3 MHz are in reasonable agreement with the simulated peak position for $Amam$, indicating that the sample might contain a small fraction of an $Amam$ phase or a closely related structural variation.
}
\label{La2ab}
\end{figure}

Figure \ref{La2ab}(a) shows the $^{139}$La(2)-NQR spectrum (3$\nu_Q$) at $T$ = 160 K ($>T_{\rm SDW}$). 
As reported previously~\cite{Yashima327}, two prominent NQR peaks from two La(2) sites are observed, labeled La(2)$_{\rm a}$ and La(2)$_{\rm b}$. Although a minimal signal from the La4310 secondary phase (approximately 4\%) is present at lower frequencies, it has a negligible effect on the overall La$_3$Ni$_2$O$_7$ spectrum or our main conclusions. 
The sharp spectral linewidths ($\le$60 kHz for La(2)$_{\rm a}$), combined with this minimal level of contamination, guarantee the microscopically high quality of the La$_3$Ni$_2$O$_{7.0}$ sample available to date.
The presence of two distinct La(2) signals cannot be explained by the ideal $Amam$ structure, which accommodates only a single La(2) site. 
Site-specific NQR parameters $\nu_Q^{\rm a,b}$ and $\eta^{\rm a,b}$ for the La(2)$_{\rm a,b}$ sites are experimentally obtained by the analysis of the resonance frequencies at $T$ = 160 K, including the $1\nu_Q$ and $2\nu_Q$ transitions (not shown), as summarized in Table \ref{NQR-Table}.
These values are compared with the NQR parameters, $\nu_Q^{\rm cal}$ and $\eta^{\rm cal}$, calculated by the DFT+$U$ method for ideal crystals of $Am2m$, $Amam$, and $Fmmm$.
The calculated peak positions for each structure are indicated in Fig. \ref{La2ab}(b). 
For the ideal $Am2m$ structure, where the La(2) site inherently splits into two inequivalent positions, the calculations indeed yield distinct NQR parameters for the two La(2)$_{\rm a,b}$ sites that are in good agreement with the experimental results.
In addition, it is noteworthy that the integrated intensity ratio between La(2)$_{\rm a}$ and La(2)$_{\rm b}$ is nearly equal, indicating that the occupancy ratio is approximately $1:1$ for the two sites.
This feature has been consistently observed in NQR studies of various La$_3$Ni$_2$O$_{7+\delta}$ crystals~\cite{Yashima327,Wang_La2PrNi2O7,Luo,Zhao_NMR,InoHPR}. 
Thus, the origin of the two La(2)$_{\rm a,b}$ sites with a 1:1 occupancy ratio can be naturally attributed to the $Am2m$ symmetry, which inherently incorporates these two inequivalent sites in equal population.
Furthermore, this $Am2m$ model also accounts for the nearly identical temperature ($T$) dependence of the nuclear spin-lattice relaxation rate ($1/T_1$) observed at the two La(2)$_{\rm a,b}$ sites~\cite{Yashima327}, since both sites are hyperfine-coupled to the same NiO$_2$ plane within the single-phase material.


Here, we reconsider the problems of the previous analyses~\cite{Yashima327,Wang_La2PrNi2O7,Luo,Zhao_NMR,InoHPR} based on the assumption of $Amam$ that contains only a single La(2) site.
To account for the additional La(2)$_{\rm b}$ signal, previous studies proposed three possible defect-related scenarios: (i) stacking faults of the La4310 phase~\cite{Wang_La2PrNi2O7,Luo}, (ii) contamination by the $Fmmm$ phase~\cite{Zhao_NMR}, and (iii) oxygen vacancies~\cite{Yashima327}.
The first scenario attributes the La(2)$_{\rm b}$ site to interfacial regions formed by stacking faults between the La4310 and La327 phases, owing to the presence of substantial amounts of La4310 impurities in their studied samples~\cite{Wang_La2PrNi2O7,Luo}.
However, this is inconsistent with the observation of a pronounced La(2)$_{\rm b}$ signal even in high-quality La327 crystals containing negligible amounts of La4310 impurities~\cite{Yashima327}.
The second scenario attributes the La(2)$_{\rm b}$ site to phase separation involving the $Fmmm$ phase~\cite{Zhao_NMR,Neutron,Fmmm-Zhang,Zhang_JSSC2026}.
However, it is difficult to explain why the integrated intensities of the La(2)$_{\rm a}$ and La(2)$_{\rm b}$ signals remain nearly equal across various samples~\cite{Yashima327,Wang_La2PrNi2O7,Luo,Zhao_NMR,InoHPR}. 
Furthermore, this scenario fails to account for the identical $T$ dependence of $1/T_1T$ as well as the same $T_{\rm SDW}$ observed at both sites~\cite{Yashima327}.
The third scenario attributes the La(2)$_{\rm b}$ to La(2) positions close to inner apical oxygen vacancies located between the two NiO$_2$ planes~\cite{Ovacancy,Yashima327}. 
However, a similar contradiction arises regarding the intensity ratio between La(2)$_{\rm a}$ and La(2)$_{\rm b}$, which should vary across samples with different oxygen concentrations.
Taken together, these inconsistencies indicate that none of the defect-based scenarios within the $Amam$ framework can consistently explain the experimental observations. 
Consequently, we deduce that the $Am2m$ symmetry represents the intrinsic crystal structure of ideal La$_3$Ni$_2$O$_{7.0}$ at ambient pressure.

\begin{figure}[!t]
\includegraphics[width=8.5cm]{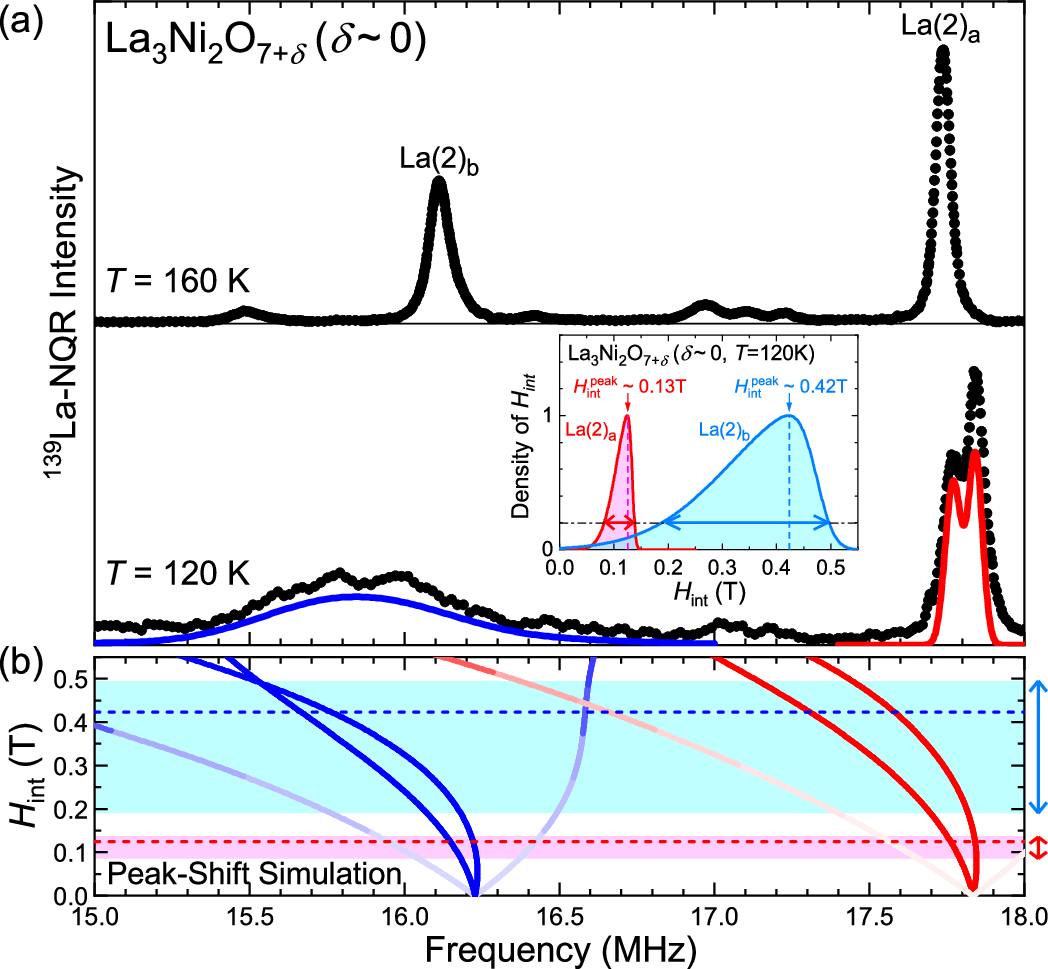}
\vspace*{-0.1cm}
\caption[]{\footnotesize (Color online) 
(a) $^{139}$La(2)-NQR spectra of the 3$\nu_Q$ transition measured across $T_{\rm SDW}$ in La$_3$Ni$_2$O$_7$.
The solid lines at $T$ = 120 K represent spectral simulations.
The inset shows the distributions of $H_{\rm int}$ used for the simulations at $T$ = 120 K, indicating their peak positions $H_{\rm int}^{\rm peak}$ and full widths at one-fifth maximum for La(2)$_{\rm a}$ (red) and La(2)$_{\rm b}$ (blue).
Note that the broader $H_{\rm int}$ distribution for La(2)$_{\rm b}$ compared to La(2)$_{\rm a}$ is roughly scaled with the larger local $M_{\rm Ni}^{\rm b}$ than $M_{\rm Ni}^{\rm a}$.
(b) Simulation of resonance peak frequencies at 3$\nu_Q$ as a function of $H_{\rm int}$ with $\theta$ = 87$^\circ$ for La(2)$_{\rm a}$ (red) and La(2)$_{\rm b}$ (blue).
The experimental spectra are reasonably reproduced by the simulations, assuming $H_{\rm int}^{\rm peak} \approx 0.13$ T for La(2)$_{\rm a}$ (red curve) and $H_{\rm int}^{\rm peak} \approx 0.42$ T for La(2)$_{\rm b}$ (blue curve). 
}
\label{Hint}
\end{figure}

\begin{figure}[tb]
\hspace*{0.2cm}
\includegraphics[width=7.5cm]{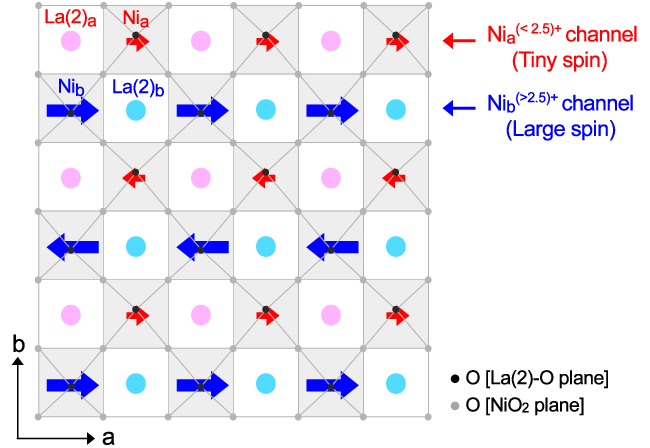}
\vspace*{-0.1cm}
\caption[]{\footnotesize (Color online)
Schematic spin configuration in the NiO$_2$ plane, showing the large-tiny spin-stripe SDW order.
The Ni spins (represented by arrows) are oriented approximately along the $a$ axis and form an alternating stripe pattern with $Q_{\rm SDW}=(0, 1/2)$ in the orthorhombic basis.
The La(2)$_{\rm a}$ and La(2)$_{\rm b}$ sites are projected onto this plane and aligned above the Ni$_{\rm a}$ and Ni$_{\rm b}$ channels that carry tiny and large $M_{\rm Ni}$, respectively.
The gray shading represents the NiO$_6$ octahedra.
Note that the single crystallographically equivalent Ni and La(2) sites in the $Amam$ and $Fmmm$ structures correspond to the uniform states without the site-distinction shown here, whereas the $Am2m$ structure features two distinct Ni sites with slightly different octahedral volumes (i.e., the Ni$_{\rm a}$O$_6$ octahedron is slightly larger than the Ni$_{\rm b}$O$_6$ octahedron).
}
\label{SDW}
\end{figure}

Next, we reconsider the NQR spectral analyses in the SDW state below $T_{\rm SDW}$ within the $Am2m$ framework, which consistently accounts for both the La(2)$_{\rm a}$ and La(2)$_{\rm b}$ spectra, whereas our previous analysis, which led us to propose a spin-spinless stripe model, was based on the assumed $Amam$ structure~\cite{Yashima327}.
Figure \ref{Hint}(a) shows the $3\nu_{\rm Q}$ spectra at La(2) measured across $T_{\rm SDW}$.
As discussed in our previous paper, the asymmetric deformation of the spectra below $T_{\rm SDW}$ excludes an SDW order that produces $H_{\rm int}$ at La(2) approximately along the $c$ axis~\cite{Yashima327}.
We attempt to simulate the spectra by varying the magnitude of $H_{\rm int}$ and the angle ($\theta$) between $H_{\rm int}$ and $V_{zz}$ at each La(2) site, taking into account that $V_{zz}$ at La(2) is tilted by about $24^\circ$ from the $c$ axis toward the $b$ axis, which is evaluated by our DFT calculations.
The red and blue curves in Fig. \ref{Hint}(b) are the peak-shift simulations at $3\nu_{\rm Q}$ for La(2)$_{\rm a}$ and La(2)$_{\rm b}$ as a function of $H_{\rm int}$ under the same $\theta$ (= $87^\circ$), respectively.
The $3\nu_{\rm Q}$ spectra in the SDW state at 120 K for La(2)$_{\rm a}$ and La(2)$_{\rm b}$ are reasonably reproduced with $(H_{\rm int}^{\rm peak}, \theta) \approx(0.13~\mathrm{T}, 87^\circ)$ and $(0.42~\mathrm{T}, 87^\circ)$, as shown by the red and blue curves in Fig. \ref{Hint}(a), respectively.
Here, we adopted the skew normal distribution $S(x, \xi, \sigma, \alpha)$ for $H_{\rm int}$ introduced in Ref. \onlinecite{Yashima327}, as shown in the inset of Fig. \ref{Hint}(a) (see Supplemental Material of Ref.~\onlinecite{Yashima327} for details).
The distribution parameters $(\xi, \sigma, \alpha)$ were set to $(0.134~\mathrm{T}, 0.04~\mathrm{T}, -6)$ for La(2)$_{\rm a}$ and $(0.474~\mathrm{T}, 0.22~\mathrm{T}, -6)$ for La(2)$_{\rm b}$, yielding $H_{\rm int}^{\rm peak} \approx 0.13~\mathrm{T}$ and $0.42~\mathrm{T}$, respectively.
Since the $Am2m$ structure intrinsically contains two alternating inequivalent Ni sites, the tiny and large $H_{\rm int}$ at La(2)$_{\rm a}$ and La(2)$_{\rm b}$ are naturally attributed to the tiny and large moments at the Ni$_{\rm a}$ and Ni$_{\rm b}$ sites, respectively, as illustrated in Fig.~\ref{SDW}. 
In this large-tiny spin-stripe model, the La(2)$_{\rm a}$ sites are aligned above the Ni$_{\rm a}$ (tiny-spin) channels along the $a$ axis, resulting in a small $H_{\rm int}$. Although each La(2)$_{\rm a}$ site is also sandwiched by two neighboring Ni$_{\rm b}$ spins (large-spin), the contributions to $H_{\rm int}$ from the Ni$_{\rm b}$ spins cancel out due to their antiparallel alignment. Conversely, the La(2)$_{\rm b}$ sites are aligned above the Ni$_{\rm b}$ (large-spin) channels, thereby experiencing a much larger $H_{\rm int}$.
Importantly, this revised magnetic structure model does not fundamentally alter our previous conclusions regarding the SDW order (single spin-spinless stripe)~\cite{Yashima327}: the Ni moments remain aligned approximately along the $a$ axis with an in-plane ordering wave vector of $Q_{\rm SDW}=(0, 1/2)$ in the orthorhombic basis, corresponding to $(1/4, 1/4)$ in the tetragonal basis, with the sole exception that the spinless sites proposed in our previous report are now reidentified as tiny-spin sites.
The large-tiny spin-stripe model has also been suggested in the recent neutron scattering~\cite{Plokhikh}, our recent $^{17}$O-NMR measurements~\cite{Lee}, and theoretical studies~\cite{LaBollita,Braz}.

Finally, we discuss the magnitude of the Ni magnetic moments ($M_{\rm Ni}$) in the large-tiny spin-stripe model based on the $Am2m$ framework.
A similar argument from our previous model~\cite{Yashima327} can be applied here. 
The magnetic moment at the large-spin Ni$_{\rm b}$ site, $M_{\rm Ni}^{\rm b}$, is estimated to be approximately $0.42$--$0.67\mu_{\rm B}$ based on previous reports~\cite{Chen_muSR,Yashima327,Plokhikh,Chen-single-neutron}.
As for La(2)$_{\rm a}$, since the $H_{\rm int}$ is approximately one-third of that at La(2)$_{\rm b}$, the magnetic moment at the tiny-spin Ni$_{\rm a}$ site, $M_{\rm Ni}^{\rm a}$, is estimated to be approximately 0.13--$0.21\mu_{\rm B}$, assuming that $H_{\rm int}$ scales approximately with $M_{\rm Ni}$.
According to the SXRD study, the NiO$_6$ octahedron at the Ni$_{\rm a}$ site has a larger volume than that at the Ni$_{\rm b}$ site, from which the valences of Ni$_{\rm a}$ and Ni$_{\rm b}$ were estimated to be +2.3 and +2.7, respectively, based on the bond-valence-sum (BVS) analysis~\cite{Misawa}. 
On this basis, they proposed that an inherent charge disproportionation exists substantially.
In contrast, our DFT+$U$ analysis yields Ni valences of +2.48 and +2.52 for Ni$_{\rm a}$ and Ni$_{\rm b}$, respectively, if the average valence of the two sites is assumed to be +2.5. 
Here, we emphasize that this calculation was performed under the same conditions that reproduces the experimental NQR parameters for the $Am2m$ structure, as discussed in Table~\ref{NQR-Table} and Fig.~\ref{La2ab}(b).
We point out that the amplitude of the valence difference obtained in our evaluation is significantly smaller than that obtained from the BVS analysis~\cite{Misawa}, although the trend in the valence contrast between the two sites is consistent. This raises an important open question: whether a large charge ordering takes place, or a minor valence contrast merely arises spontaneously due to the structural inequivalence between the two Ni sites.
Assuming the localized-moment picture, $M_{\rm Ni}^{\rm a} > M_{\rm Ni}^{\rm b}$ would be naturally expected, since the Ni$_{\rm a}$ site (+2.48 or +2.3~\cite{Misawa}) has a larger electron occupancy than the Ni$_{\rm b}$ site (+2.52 or +2.7~\cite{Misawa}). 
This is inconsistent with our experimental results that $M_{\rm Ni}^{\rm b} \gtrsim 3 M_{\rm Ni}^{\rm a}$, indicating that the SDW order in La327 is driven not by a localized-moment mechanism but by itinerant Fermi-surface nesting.


In summary, we have reexamined the crystal structure and spin configuration of La$_3$Ni$_2$O$_7$ microscopically via $^{139}$La(2)-NQR spectroscopy. 
The experimentally obtained NQR parameters for the two distinct La(2)$_{\rm a}$ and La(2)$_{\rm b}$ are naturally reproduced by DFT+$U$ calculations based on $Am2m$ without any special assumptions. 
This finding resolves the long-standing inconsistencies in previous NQR analyses based on the Amam~\cite{Yashima327,Wang_La2PrNi2O7,Zhao_NMR,Luo} by demonstrating that both La(2) sites are intrinsic to the $Am2m$ structure. 
Crucially, the SDW spin structure is identified as a large-tiny spin-stripe order within the $Am2m$ framework, which naturally accommodates two inequivalent Ni sites with large and tiny $M_{\rm Ni}$.
Furthermore, the inconsistency between the valence contrast and $M_{\rm Ni}$ for the two Ni sites indicates that the SDW order at ambient pressure is driven by the Fermi-surface nesting.
These findings establish the itinerant nature of the SDW order at ambient pressure, providing a basis for understanding its relationship to the emergence of high-$T_c$ superconductivity under high pressure.


\vspace*{0.3cm}
{\footnotesize 
We thank R. Misawa and M. Hirschberger for providing us with structural data, and K. Kuroki for valuable comments.
This work was supported by JSPS KAKENHI (JP25K00959, JP26K22290, and JP24K01333).
M. K. was supported by a JSPS Fellowship (JP25KJ1758), and H. S. and Y. T. by the World Premier International Research Center Initiative (WPI), MEXT, Japan.
}


\end{document}